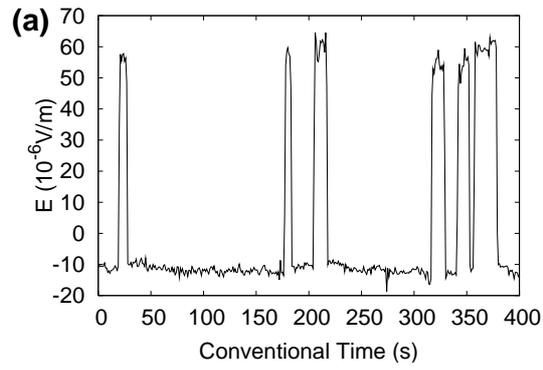

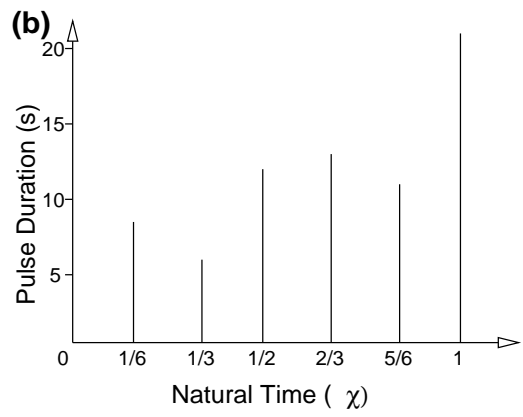

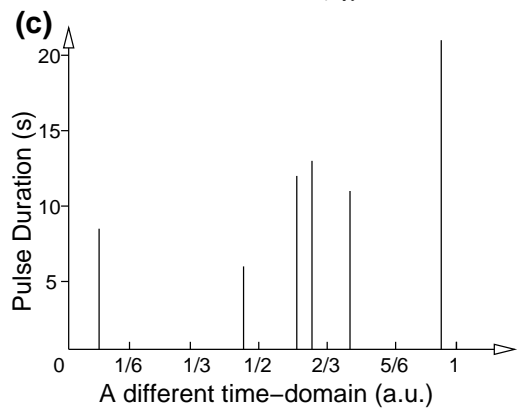

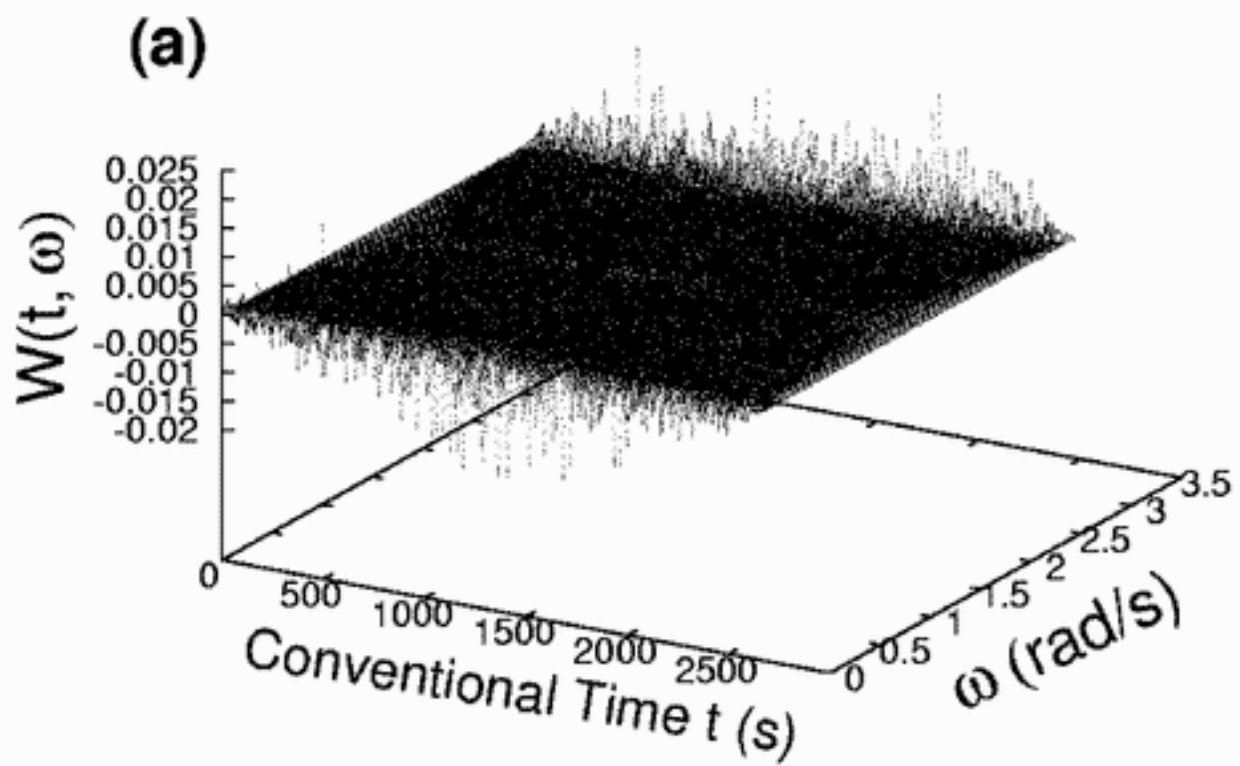

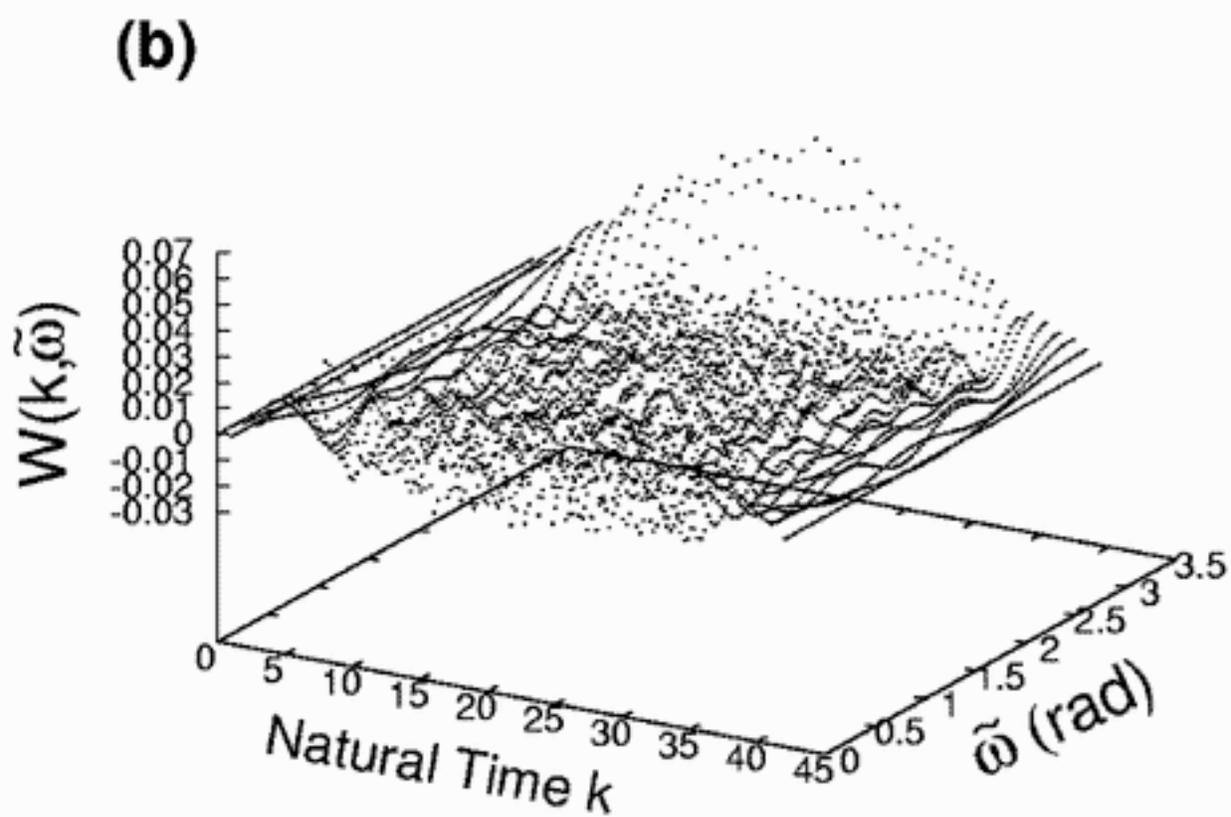

# Optimality of Natural Time Representation of Complex Time Series

Sumiyoshi Abe[1], N. V. Sarlis[2], E. S. Skordas[2,3], H. Tanaka[2,4], and P. A. Varotsos[2,3]

[1]*Institute of Physics, University of Tsukuba, Ibaraki 305-8571, Japan*

[2]*Solid State Section, Physics Department, University of Athens,*

*Panepistimiopolis, Zografos, Athens, 157 84, Greece*

[3]*Solid Earth Physics Institute, Physics Department, University of Athens,*

*Panepistimiopolis, Zografos, Athens, 157 84, Greece*

[4]*Earthquake Prediction Research Center, Tokai University,*

*Shizuoka 424-8610, Japan*

**Abstract**

The concept of natural time turned out to be useful in revealing dynamical features behind complex time series including electrocardiograms, ionic current fluctuations of membrane channels, seismic electric signals, and seismic event correlation. However, the origin of this empirical usefulness is yet to be clarified. Here, it is shown that this time domain is in fact optimal for enhancing the signals in time-frequency space by employing the Wigner function and measuring its localization property.

Characterization of complex time series and prediction of catastrophic events have always been of general common interest in biology, earth science, and physics. In traditional analysis, no attention has been paid to the concept and role of time itself and the possibility of introducing its reparametrization. In recent works [1-7], it has been shown that novel dynamical features hidden behind time series can emerge if they are represented in terms of "natural time". Natural time, $\chi$, is defined [1,2] by ascribing to the $k$th pulse (once the initial pulse identified) the value $\chi_k = k/N$, where N is the total number of pulses considered. Using this new reparametrization in time series analysis, it has been successful in discriminating sudden cardiac death individuals from healthy humans through analysis of their electrocardiograms [5], discriminating Seismic Electric Signals (SES) activities (i.e., series of electrical pulses detected before earthquakes [8-10]) from irrelevant background noise [3,4], and manifesting aging and scaling properties in seismic event correlation [6,7]. We emphasize that these results could not be obtained if the analyses were carried out in the conventional time domain. The most important point regarding natural time may be that it enables to follow dynamical evolution of a system and identify when it enters into a critical stage. Therefore, it can play a major role in predicting an impending catastrophic events such as a strong earthquake occurrence [10] and sudden cardiac death [11]. However, the question remains to be solved why natural time exhibits more advantages than conventional time.

In this paper, we address ourselves to the problem of optimality of the natural time representation of time series resulting from complex systems that may contain catastrophic events. For this purpose, first we study the structures of the time-frequency representations [12] of the signals by employing the Wigner function [13] to compare the natural time representation with the ones, either in conventional time or other possible reparametrizations. We shall see that significant enhancement of the signal is observed in the time-frequency space if natural time is used, in marked contrast to other time domains. To quantify this localization property, we examine the generalized entropic measure proposed by Tsallis [14], which has been widely discussed in the studies of complex dynamical systems. In time series analysis, it is desired to reduce uncertainty and extract signal information as much as possible. Consequently, the most useful time domain should maximize the information measure, and hence minimize the entropy. We find that this can statistically be ascertained in natural time, by investigating a multitude of different time domains.

Consider a signal $\{x(t)\}$ represented in conventional time, $t$. The normalized time-frequency Wigner function associated with it is defined by

$$W(t,\omega) = A \int d\tau e^{-i\omega\tau} x(t-\tau/2) x(t+\tau/2), \qquad (1)$$

where $A = [\pi \int dt \ x^2(t)]^{-1}$ is the normalization constant and ω is the frequency. Numerically, it is necessary to discretize and make finite both time and frequency, and the integral has to be replaced by a sum. In the natural time representation, the signal $\{x(t)\}$ is substituted by the sequence of pairs $\{\chi_k, Q_k\}$, where $Q_k$ stands for the duration (in conventional time) of the $k$th pulse [1-4]. To make comparison of the natural time analysis with Eq. (1), it is convenient to rescale $\chi_k$ by $N\chi_k$, which is precisely the pulse number, $k \equiv t_k$. The quantity, $Q_k$, has a clear meaning for dichotomous time series (Fig. 1), whereas for nondichotomous time series, threshold should be appropriately put (e.g., the mean value plus half of the standard deviation) to transform it to a dichotomous one. The normalized Wigner function associated with $Q_k$ is now given as follows:

$$W(k,\omega) = B\sum_{i=0}^{N-1} Q_{k-i} Q_{k+i} \cos[\widetilde{\omega}(t_{k+i} - t_{k-i})], \qquad (2)$$

where $B = [\pi \sum_{k=1}^{N} Q_k^2]^{-1}$ stands for the normalization constant and $\widetilde{\omega}$ is the dimensionless "frequency" (see the later comment). In the sum, $Q_k$ with $k \leq 0$ and $k > N$ should be set equal to zero. It is noted that Eq. (2) is a discrete version of the continuous Wigner function in Eq. (1) and unlike the ordinary definition the transformation in Eq. (2) is not orthogonal, in general.

In Fig. 2, we present the plots of the Wigner functions in the time-frequency spaces with conventional time and natural time. Remarkably, significant enhancement of the signal is observed in the latter case, with the scale of enhancement being about 10 times. In contrast to a moderate profile in the conventional time representation, a localized structure emerges in natural time.

In the natural time domain, the time difference between two consecutive pulses (i.e., inter-occurrence time) is equally spaced and dimensionless, and is here taken to be unity: $t_{k+1} - t_k = 1$. However, for comparison, later we will consider various time domains, in which the occurrence time $t_k = Nu_k$ in Eq. (2) is made random. The conventional time representation is characterized by a constant time increment $\Delta t$ (e.g., 1 sec), and the occurrence of the ith event is at $t_i = i\Delta t$. Differences between three time domains are shown in Fig. 1. To generate the random time domains artificially, we randomize $u_k$ by making use of the uniform distribution defined in the interval (0,1) so that the average inter-occurrence time is again unity. Performing Monte-Carlo simulation, we have constructed more than 1000 different time domains and integrated over ω ($\tilde{\omega}$) over 0 to π [rad/sec] ([rad]), which can cover the regimes of interest (recall that when $t_k = k$, $W(k, \omega + \pi) = W(k, \omega)$).

To quantify the degrees of disorder in the time-frequency spaces with various time domains, we employ as mentioned the Tsallis entropy [14] defined by

$$S_q = \frac{1}{1-q}(\int d\mu W^q - 1),  \qquad (3)$$

where $\int d\mu$ is the collective notation for integral and sum over the time-frequency space and $q$ is the positive entropic index. In the limit $q \to 1$, this quantity tends to the form of the Boltzmann-Gibbs-Shannon entropy $S = -\int d\mu W \ln W$. This limit cannot however be taken, since the Wigner function is a pseudo-distribution and takes negative values, in general. $S_q$ is however well defined if $q$ is even. Taking into account the fact that the negative contributions are not significant (see Fig. 2), we propose to use the value

$$q = 2, \qquad (4)$$

which, using Eqs.(2) and (3), results in:

$$S_2 = 1 - \frac{1}{2\pi}\left\{\frac{\sum_{k=1}^{N}\sum_{l=0}^{N-1}\sum_{l'=0}^{N-1} Q_{k-l}Q_{k+l}Q_{k-l'}Q_{k+l'}\left(\delta_{l+l',0} + \frac{\sin[\pi(t_{k+l}-t_{k-l}+t_{k+l'}-t_{k-l'})]}{\pi(t_{k+l}-t_{k-l}+t_{k+l'}-t_{k-l'})} + \delta_{l,l'} + \frac{\sin[\pi(t_{k+l}-t_{k-l}-t_{k+l'}+t_{k-l'})]}{\pi(t_{k+l}-t_{k-l}-t_{k+l'}+t_{k-l'})}\right)}{\left[\sum_{k=1}^{N}\left(Q_k^2 + \sum_{l=1}^{N-1} Q_{k-l}Q_{k+l}\frac{\sin[\pi(t_{k+l}-t_{k-l})]}{\pi(t_{k+l}-t_{k-l})}\right)\right]^2}\right\} \qquad (5)$$

To examine how the natural time representation is superior to other ones, we have made comparison of the values of $S_2$ for 10 different time series[4] of electric signals (see Fig. 3): 4 SES activities and 6 "artificial" noises (i.e., noises emitted from nearby electrical sources). The results are shown in Table 1 in which we give the values of $p(S_2 < S_{nat})$,.i.e., the probability that $S_2$ calculated with a time domain different than the natural time domain to be smaller than the value $S_2^{nat}$ calculated with natural time. An inspection of this Table shows that among the signals investigated only two, i.e., A and n6, have a considerable probability $p(S_2 < S_2^{nat})$, i.e., ~28.5 and 26%, respectively. This can be attributed to the small number of pulses (N≈40) of these signals for the following reason: In Fig. 4 we present the dependence of $p(S_2 < S_2^{nat})$ versus the number of pulses for the simplified example of all $Q_k$=1; this figure shows that, $p(S_2 < S_2^{nat})$ decreases upon increasing N, starting from ~36% at N~50. In other words Table 1 reveals that, for signals with a reasonable number of pulses, e.g., larger than $2 \times 10^2$, the quantity $S_2^{nat}$, in fact, tends to be minimum compared to those of other representations attempted. In addition, it is mentioned that $S_2^{nat}$ is also appreciably smaller than $S_2$ in conventional time (see Fig. 2).

In conclusion, we have studied if natural time yields an optimal representation for enhancing the signals in the time-frequency space by employing the Wigner function and measuring its localization property by means of the Tsallis entropy. For this purpose, we have compared the values of the entropy for various observed time series represented in a multitude of different time domains. We have found that the entropy is highly likely to be minimum for natural time, implying the least uncertainty in the time-frequency space. This explains why dynamical evolutions of diverse systems can be better described in the natural time domain, in particular when systems approach to critical state.

S.A. would like to thank Physics Department of the University of Athens for hospitality extended to him. He was also supported in part by the Grant-in Aid for Scientific Research of Japan Society for the Promotion of Science.[1]    P. Varotsos, N. Sarlis, and E. Skordas, Practica of Athens Academy **76**, 294 (2001).

[2]    P. A. Varotsos, N. V. Sarlis, and E. F. Skordas, Phys. Rev. E **66**, 011902 (2002).

[3]    P. A. Varotsos, N. V. Sarlis, and E. F. Skordas, Phys. Rev. E **67**, 021109 (2003).

[4]    P. A. Varotsos, N. V. Sarlis, and E. F. Skordas, Phys. Rev. E **68**, 031106 (2003).

[5]    P. A. Varotsos, N. V. Sarlis, E. F. Skordas, and M. S. Lazaridou, Phys. Rev. E **70**, 011106 (2004).

[6]    S. Abe and N. Suzuki, Physica A **332**, 533 (2004).

[7]    U. Tirnakli and S. Abe, Phys. Rev. E **70**, 056120 (2004).

[8]    P. Varotsos, K. Alexopoulos, K. Nomicos, and M. Lazaridou, Nature (London) **322**, 120 (1986).

[9]    S. Uyeda, T. Nagao, Y. Orihara, T. Yamaguchi, and I. Takahashi, Proc. Natl. Acad. Sci. U.S.A. **97**, 4561 (2000).

[10]    P. Varotsos, *The Physics of Seismic Electric Signals* (TERRAPUB, Tokyo, in press).

[11]    D. Graham-Rowe, NewScientist (3 April, 2004), p. 10.

[12]    L. Cohen, *Time-Frequency Analysis: Theory and Applications* (Prentice-Hall, Upper Saddle River NJ, 1994).

[13]    E. Wigner, Phys. Rev. **40**, 749 (1932).7

# Figure Captions

**Fig. 1** An example of observed time series of SES activity represented in (a) conventional time, (b) natural time, and (c) a randomly generated time.

**Fig. 2** The plots of the Wigner functions of the SES activity "n6" in Fig. 3 given Below in (a) the conventional time domain and (b) the natural time domain. Significant enhancement of the signal is recognized in the natural time domain. Note that, instead of $\chi_k$, $N\chi_k = k$ is used (see the text). $\omega$ has the unit [rad/sec], whereas $\tilde{\omega}$ has [rad].

**Fig. 3** Excerpts of 4 SES activities, labeled K1, K2, A, U and 6 "artificial" noises, labeled n1-n6, in arbitrary scales.

**Fig. 4** The values of $p(S_2 < S_2^{nat})$ versus the number of pulses for the simple example of a time series consisting of pulses with all $Q_k=1$.

**Table I** The number of N pulses and the values of $p(S_2 < S_2^{nat})$ for the 10 electric signals analyzed. The estimation error is at the most 1.6%.

**Table I**

| Signal | $N$ | $p(S_2 < S_2^{nat})$ |
|---|---|---|
| | | % |
| K1 | 312 | 3.7 |
| K2 | 141 | 6.9 |
| A | 43 | 28.5 |
| U | 80 | 8.1 |
| n6 | 42 | 26.0 |
| n5 | 432 | 2.8 |
| n4 | 396 | 1.6 |
| n3 | 259 | 2.7 |
| n2 | 1080 | <0.1 |
| n1 | 216 | 5.7 |

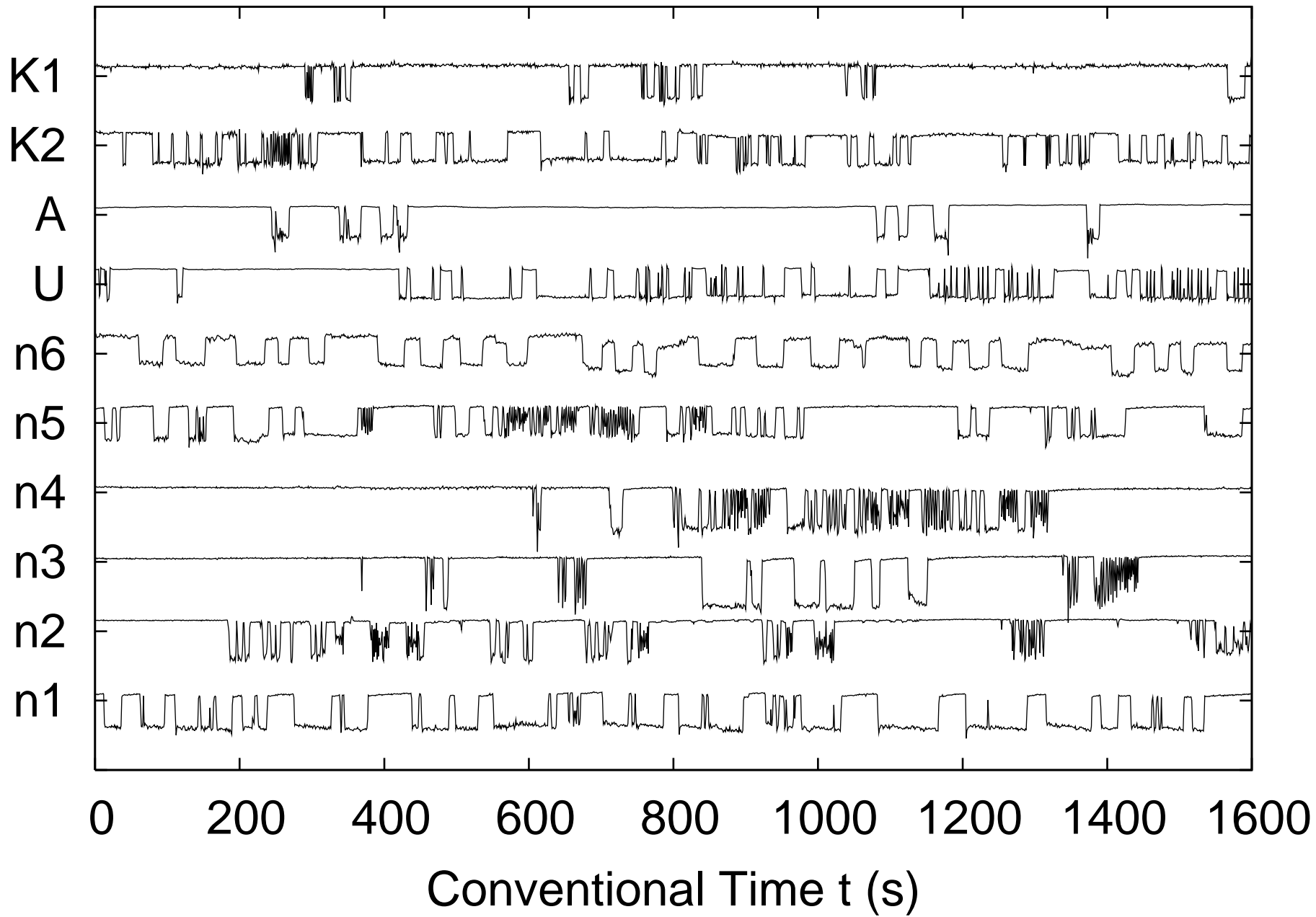

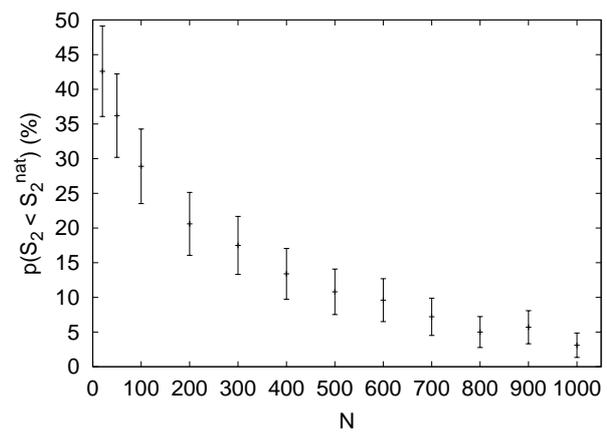